\documentclass{article}

\usepackage{PRIMEarxiv}

\usepackage[utf8]{inputenc} 
\usepackage[T1]{fontenc}    
\usepackage{hyperref}       
\usepackage{url}            
\usepackage{booktabs}       
\usepackage{amsfonts}       
\usepackage{nicefrac}       
\usepackage{microtype}      
\usepackage{lipsum}
\usepackage{fancyhdr}       
\usepackage{graphicx}       
\graphicspath{{media/}}     

\title{DeepClouds.ai: Deep learning enabled computationally cheap direct numerical simulations
}

\author{
  Moumita Bhowmik \\
  Indian Institute of Tropical Meteorology \\
  Pune, India\\
  \texttt{moumita.bhowmik@tropmet.res.in} \\
   \And
  Manmeet Singh \\
  Indian Institute of Tropical Meteorology \\
  Pune, India\\
  \texttt{manmeet.cat@tropmet.res.in} \\
  \And
  Suryachandra Rao \\
  Indian Institute of Tropical Meteorology \\
  Pune, India\\
  \texttt{surya@tropmet.res.in} \\
  \And
  Souvik Paul \\
  Hewlett Packard Enterprise \\
  Pune, India\\
  \texttt{souvik4389@hotmail.com} \\
}

\begin{document}
\maketitle

\begin{abstract}
Simulation of turbulent flows, especially at the edges of clouds in the atmosphere, is an inherently challenging task. Hitherto, the best possible computational method to perform such experiments is the Direct Numerical Simulation (DNS). DNS involves solving non-linear partial differential equations for
fluid flows, also known as Navier-Stokes equations, on discretized grid boxes in a three-dimensional space. It is a valuable paradigm that has guided the numerical weather prediction models to compute rainfall formation. However, DNS cannot be performed for large domains of practical utility to the weather forecast community. Here, we introduce DeepClouds.ai, a 3D-UNET that simulates the outputs of a rising cloud DNS experiment. 
The problem of increasing the domain size in DNS is addressed by mapping an inner 3D cube to the complete 3D cube from the output of the DNS discretized grid simulation. Our approach effectively captures turbulent flow dynamics without having to solve the complex dynamical core. The baseline shows that the deep learning-based simulation is comparable to the partial-differential equation-based model as measured by various score metrics. This framework can be used to further the science of turbulence and cloud flows by enabling simulations over large physical domains in the atmosphere. It would lead to cascading societal benefits by improved weather predictions via advanced parameterization schemes.
\end{abstract}

\section{Introduction}
\label{sec:intro}
One of the cornerstones of atmospheric science is the
study of cloud microphysics. Cloud prediction is crucial
for understanding atmospheric turbulence and climate
model uncertainties. Almost two-thirds of the Earth’s surface is covered by clouds \cite{Baker-97}. By reflecting solar radiation, transporting, and releasing energy, they have an impact on the Earth-atmosphere system. They also play a crucial role in regulating the atmospheric hydrological cycle, modulating atmospheric chemistry, and impacting extreme weather
events through atmospheric moisture redistribution. These phenomena have their origins in cloud microphysical processes, which have been investigated extensively in recent decades  \cite{Vail-01, Vaill-02, Pruppacher-10, Rogers-89, Lanotte-09, Sardina-15,Gotoh-16,Yang-18}. The microphysical characteristics of cloud droplets are strongly influenced by the turbulent flow surrounding them. Direct numerical simulation (DNS) is a straightforward approach for disentangling many feedbacks between small-scale turbulence and droplet dynamics by resolving the turbulence down to the smallest eddy size. Three-dimensional (3D) DNS combines the Eulerian description of turbulent flow with the Lagrangian development of an ensemble of cloud water droplets \cite{Kum-Jan-Sch-Sha-12}.


\subsection{Critical analysis}
DNS can directly solve Navier-Stokes equations without any turbulence model. However, it has some significant weaknesses, including the need for a very fine grid in DNS to capture the small-scale turbulent flow and the generation of big data for turbulent flow. As a result of these factors, DNS is a computationally expensive technique \cite{Thomas-20,Kumar-18}.
In a nutshell, due to existing computational resource constraints, DNS is not a feasible option for numerical weather prediction and climate modelling \cite{Pope-00, Speziale-91, Voller-02}. Thus, there exists a gap between the smallest scale cloud microphysics and large scale atmospheric flow \cite{Baker-08}. These limitations necessitate the need for alternate approaches to simulate small-scale flows in the atmosphere.

\subsection{Deep learning for DNS}
Deep learning has revolutionized various fields, particularly geosciences \cite{Brunton-20, Pandey-20, reichstein2019deep} within the framework of computer vision. DNS typically involves computations over three-dimensional grid boxes similar to the 3D images in the computer vision community. Recently, a number of studies have attempted the use of deep neural networks for solving turbulent flows in fluids \cite{Bhatnagar-19,Thuerey-20,Thuerey-21,Back-19,Xie-19,Xie-20,Um-20,Kochkov-21,Maulik-19,Melado-21,Cheng-19} . Introducing deep learning in turbulence will help understand the characteristics of clouds surrounding flow patterns. The knowledge of cloud flow is highly essential for the development activities in the atmospheric research community. Accurate representation of the smallest-scale microphysical processes in weather prediction models by deep learning would enable furthering the skill of numerical weather prediction (Fig. 1). Computationally cheap DNS offered by deep learning would aid the development of better parameterization techniques.


%
\subsection{Aim $\textbf{\&}$ contribution}
%
The purpose of this study is to enable the modelling of small-scale processes across broad domains. Solving Navier-Stokes equations with small-scale processes at spatial scales applicable to the weather and climate science community is not feasible at present due to the computational costs of DNS. We introduce DeepClouds.ai, which can perform the mapping of a smaller grid box(A) within the DNS domain to a larger region (B) which encapsulates the smaller region(A). Thus, A lies within B, and our deep learning model attempts to estimate B from A. The mapping of DNS outputs from a smaller three-dimensional cube to a larger three-dimensional cube is a difficult undertaking. 
\par 
Previous studies \cite{Bhatnagar-19,Thuerey-20,Thuerey-21,Back-19,Xie-19,Xie-20,Um-20,Kochkov-21,Maulik-19,Melado-21,Cheng-19} have attempted to simulate turbulence using data-driven approaches. However, none of them uses a three-dimensional representation of the flow as it occurs in reality. Further, none of the existing implementations have attempted the complex, nonlinear, and computationally difficult task of simulating three-dimensional atmospheric cloud flows. We use a three-dimensional convolutional neural network, known as a 3D-UNET, to simulate cloud flows over 2x domain from a region of size 1x. This is a test case of our newly built framework, which can be applied to different atmospheric simulations. 
Because our work is a one-of-a-kind attempt in this direction, it can be used as a baseline to benchmark datasets created in the future using DeepClouds.ai. 
\begin{figure*}
  \centering
  \includegraphics[width=0.6\linewidth]{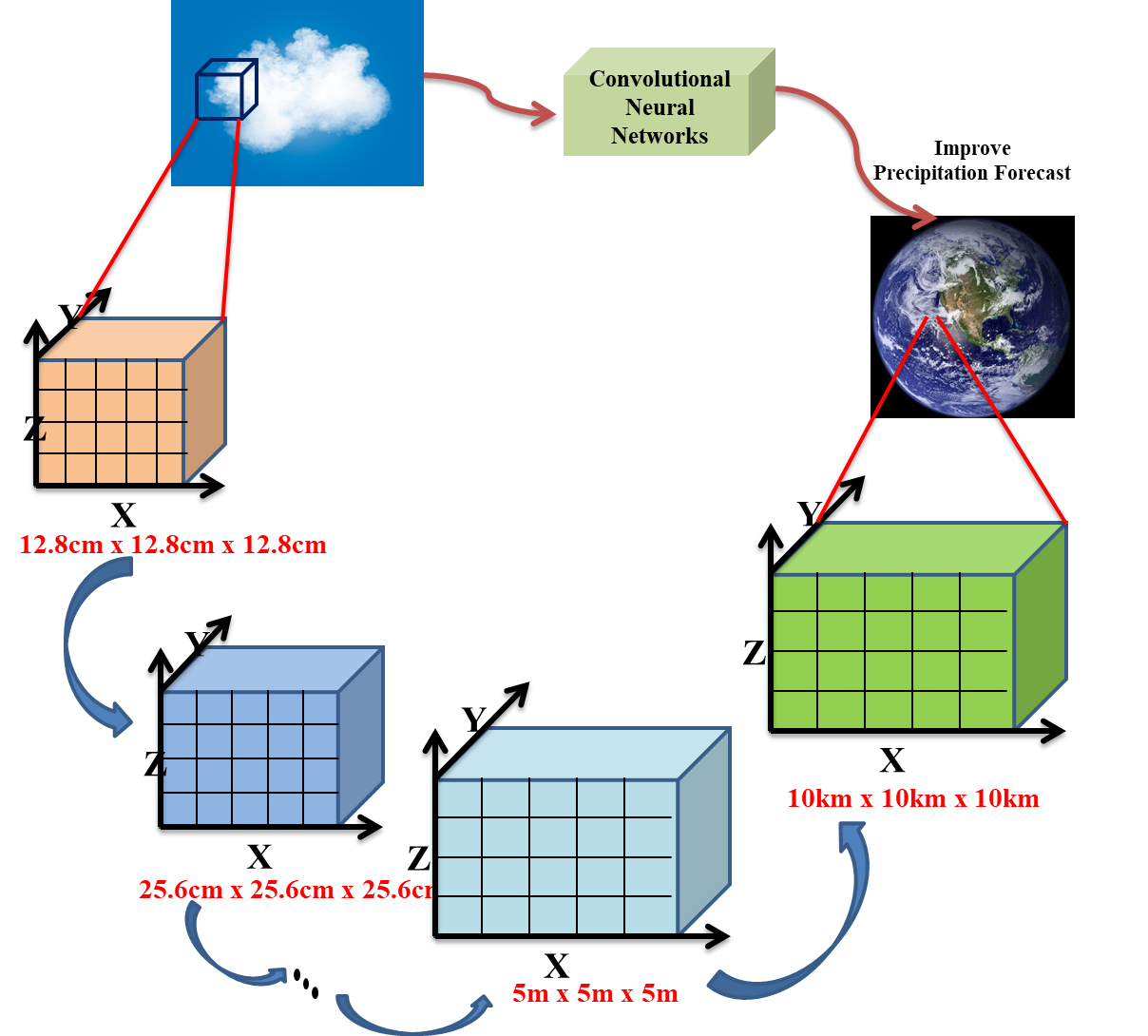}

  \caption{Schematic showing the recursive simulations of large DNS domains using DeepClouds.ai. The experiments over larger regions with small scale micro-physics would improve precipitation forecasts in the numerical weather prediction models.}
  \label{fig-1}
\end{figure*}
%

\section{Model Description}
\subsection{Flow Equations}
In this section, we describe the fluid flow equations for turbulent flow fields that form the core of DNS simulations used for training the deep learning based model. The basic equations for turbulent flow fields are as follows:
\begin{equation}
\nabla . {\bf U} = 0, ~\label{eq1}
\end{equation}
\begin{eqnarray}
\frac{\partial {\bf U}}{\partial t} + ({\bf U}.\nabla){\bf U}
= - \frac{1}{\rho_0} \nabla p 
&+&\nu \nabla^2 {\bf U} \\ \nonumber 
&&
+ B e_z + f_{LS}(x, t), ~~~\label{eq2}
\end{eqnarray}
\begin{equation}
\frac{\partial {\bf T}}{\partial t} + {\bf U}.\nabla {\bf T} = k \nabla^2 {\bf T}+ \frac{L}{c_p}C_d - \frac{g_z}{c_p}w_z, ~\label{eq3}
\end{equation}
\begin{equation}
\frac{\partial q_v}{\partial t} + {\bf U}.\nabla q_v = D \nabla^2 q_v - C_d. ~\label{eq4}
\end{equation}
Here, $\textbf{U}$ denotes flow velocity, $\textbf{T}$ is flow temperature, and ${q_v}$ is vapor mixing ratio. $B$ and $f_{LS}$ denote buoyancy term, and forcing term, respectively. 
$L$ is latent heat of evaporation, $C_p$ is specific heat, and $C_d$ is the condensation rate. A detailed description about these parameters are given in \cite{Kum-Jan-Sch-Sha-12,Kum-Sch-Sha-14}.

Pressure variation ($\textbf{P}$) is calculated from Hydrostatic equation:
\begin{equation}
\frac{d{\bf P}}{dt} = -\frac{g {\bf P} w_z}{R_t {\bf T}} ~\label{eq5}
\end{equation}
Here, $R_t$ is specific gas constant of moist air, and the vector $\textbf{g} = (0,0,-g)$ includes the gravitational acceleration $g (= 9.81 m/s^2)$. $w_z (= 0.5 m/s)$ is a constant upward velocity of rising domain.

The condensation rate ($C_d$) per unit mass inside each grid cell is determined by,
\begin{equation}
C_d(x,t) = \frac{1}{m_a} \frac{dm_l}{dt} = \frac{4\pi \rho_l}{\rho_0 a^3} \sum_{\beta = 1}^\Delta S(X_\beta,t)r(\textbf{X},t) ~\label{eq6}
\end{equation}
Here, $m_a$ is the mass of air per grid cell, the sum collects the droplets inside each grid cells of size $a^3$ that surround the grid point $x$ and $S$ is supersaturation. More details  of (Equation 6)  can be found  in \cite{Kum-Jan-Sch-Sha-12,Kum-Sch-Sha-14}.
\begin{table}[htbp]
%
\caption{List of constants, reference values and initial simulation parameters}
\begin{tabular}{|c|c|} 

\hline
 Domain Size & $(25.6 cm)^3$ \\ \hline
 Number of Grids & 256 $\times$ 256 $\times$ 256    \\ \hline
 Droplets number concentration & 343 $(cm)^3$    \\ \hline
 Initial Temperature & 14.3$^0$ C    \\ \hline
 Initial Pressure & 765 hPa    \\ \hline
 Relative Humidity of environment & 65$\%$ \\ \hline
 Initial Height  &  500 m \\ \hline
 Taylor microscale Reynolds number ($Re_\lambda$) & 59 \\ \hline
 Dissipation rate ($\epsilon$) & 1.5 $\times$ $10^{-5} m^2s^{-1}$\\ \hline
 Kolmogrove Time ($\tau_\eta$) & 122 ms \\ \hline
\end{tabular}
\label{tab1}
\end{table}
\subsection{Droplets Equations}
The liquid water component is modeled as a Lagrangian ensemble of $N_p$ pointlike droplets in the DNS domain. The droplet motion is governed by drag force and gravity, with the form:

\begin{equation}
\frac{d{\bf X}(t)}{dt} = {\bf V}({\bf X}, t), ~\label{eq7}
\end{equation}
\begin{equation}
\frac{d{\bf V}({\bf X},t)}{dt} = \frac{{\bf U}({\bf X},t) - {\bf V}({\bf X}, t)}{\tau_p} + {\bf g}, ~\label{eq8}
\end{equation}
\begin{eqnarray}
\frac{dr({\bf X},t)}{dt} &=& \frac{K}{r({\bf X},t) + \xi} (S({\bf X},t) + 1- (1+ \\ \nonumber 
&&\frac{a {r_d}^3}{{r({\bf X},t)}^3 + b {r_d}^3)}) exp(\frac{B}{r({\bf X},t)})), ~\label{eq9}
\end{eqnarray}
where, $U(x,t)$ is the flow velocity at droplet's position $X(t)$, $V(X,t)$ is droplet's velocity, $r({\bf X},t)$ is the droplet's radius, and $\tau_p$ ($= 2\rho_l r^2/(9 \rho_o \nu)$ is a finite particle response time.
A detail description about a, b, $K$, $\xi$, $r_d$ and $B$ can be found in appendix B of \cite{Kor-95}.   
\subsection{Initial Setup}
At the beginning of the DNS, cloud droplets of different sizes, containing dry aerosols are placed randomly in one-third of the domain, and the rest of the domain is filled with clear air as shown in (Fig. 2). These droplets can move freely with turbulent flow and always remain inside the domain. In other words, the domain is treated as adiabatic. The domain starts rising at about 500m from the cloud base. A 3D box of size $(25.6cm)^3$ is used for the direct numerical simulation. The flow inside the domain is periodic in all three (x,y,z)-directions. The details of the other initial parameters are given in Table 1. 
%

%
\begin{figure}[htbp]
\centerline{\includegraphics[height=8cm,width=9cm]{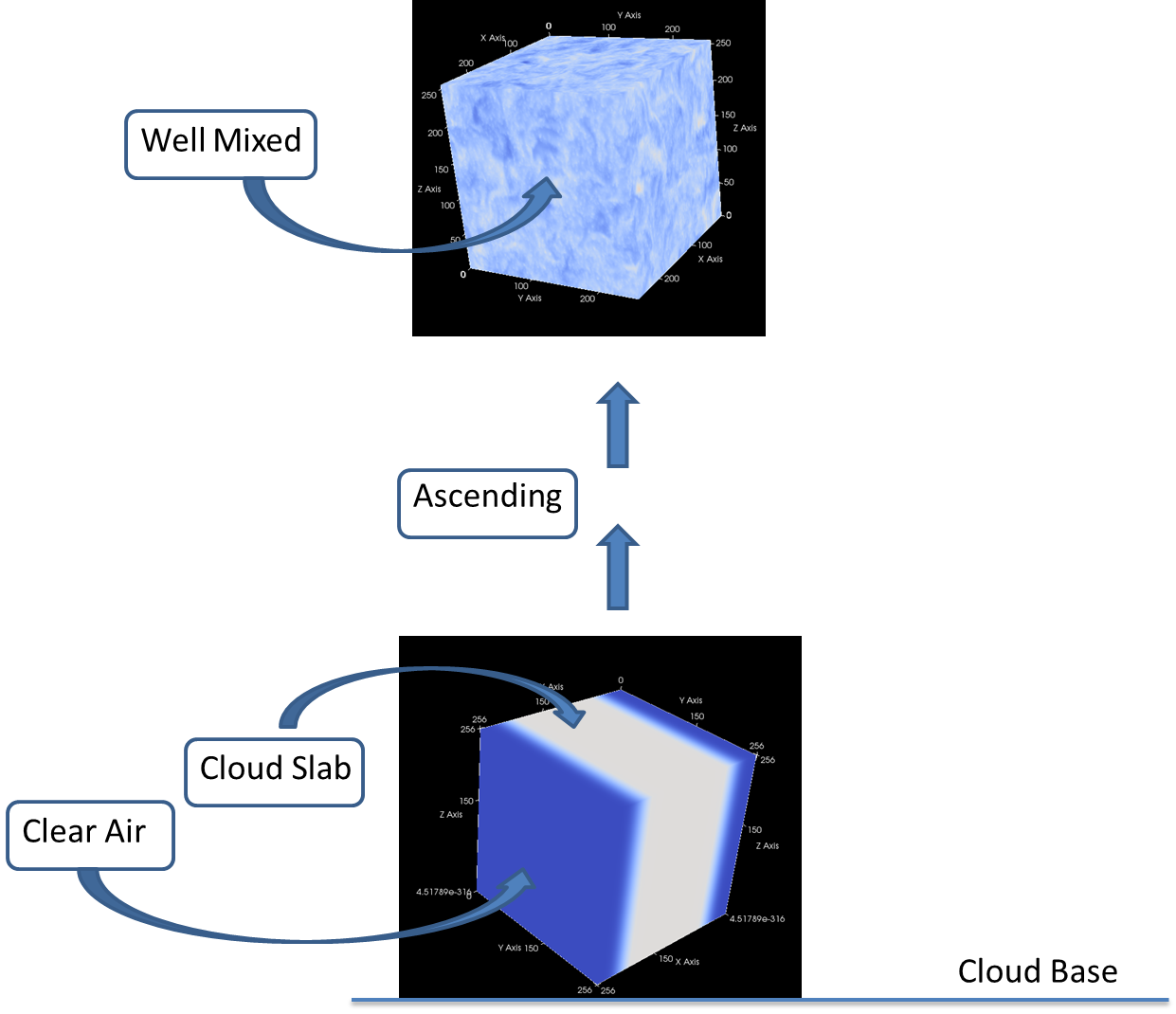}}
\caption{Conceptual diagram of rising cloud domain simulated by the Eulearian-Langrangian DNS.}
\label{fig-2}
\end{figure} 
%
\begin{figure*}
  \centering
  \includegraphics[width=0.8\linewidth]{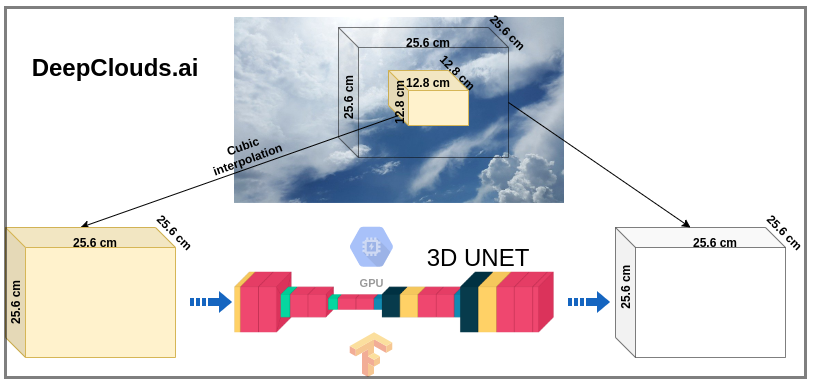}

  \caption{Schematic of data-driven simulations of clear-cloudy turbulent flows. The inner 3D cube of a direct numerical simulation model is mapped to the larger 3D cube using a UNET algorithm. The cartoon shows an imaginative view of the clear-cloudy turbulent flow simulation that is used as a test case in this study.}
  \label{fig:short}
\end{figure*}

\begin{figure}[htbp]
\centerline{\includegraphics[height=6cm,width=9cm]{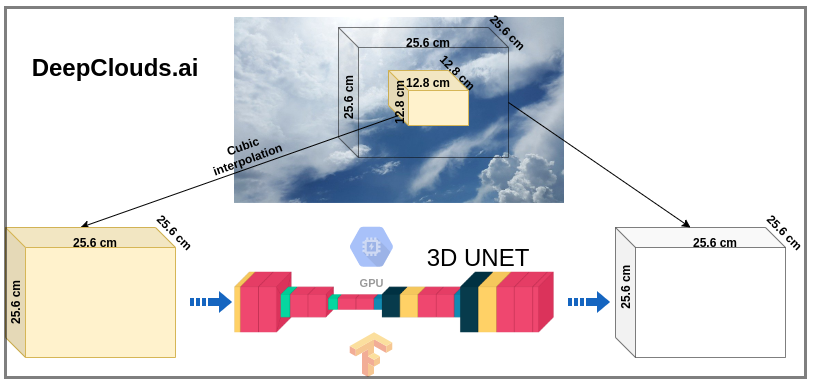}}
\caption{Evolution of mean squared error loss while training DeepClouds.ai.}.
\label{fig-33}
\end{figure} 

\section{The dataset}
We generate a labelled training dataset that consists of
the precursor as a smaller 3D cube extracted from a DNS
simulation. The label is a larger domain at the same time
step in the DNS experiment.

\subsection{DNS simulations for generating labelled data}
We ran the DNS simulation corresponding to a 256 x 256 x 256 discretized cube over a domain of size $(25.6 cm)^3$ for 600 seconds. The number of CPU processors required to perform the DNS simulation equals 512. It took a cumulative simulation time of around 30 hours. DNS is a fully parallel model and was run on an HPC system having computational capacity of 4.0 PFlops. It includes 3315 dual-socket compute nodes, each with Intel Xeon Broadwell E52695 v4 18-core processors (2.10GHz). Pratyush is a Cray XC40 supercomputer with liquid cooling. This HPC system has a total memory capacity of 414 TB, each node has 128GB DDR4 (2667 MHz) memory. All nodes are interconnected using the high speed, low latency Cray Aries Network with Dragonfly network topology, which uses per packet adaptive routing.

The DNS model is integrated with a time-step of 0.0005 seconds, and the data for turbulent flow inside the rising domain is generated every 0.2 seconds. This dataset corresponds to instantaneous fields comprising four variables. These fields are the velocity-(u,v,w) components of the flow fields and vapor mixing ratio ($q_v$) at every grid points of the 3D box. We archive a single file for each unique time step of the model because the size of DNS data is enormous, comprising of big 3D matrices for the four variables. Thus, the total number of time steps for extracting DNS output is equal to the number of saved files. The number of samples for the training datasets comprises the total number of files saved for the DNS simulation. Hence, there are 3000 samples consisting of 3D cubical matrices available from DNS simulation for generating labelled datasets. 

\section{Methodology}

 From the dataset generated by DNS simulation, precursors or inputs corresponding to the wind and mixing ratio on the inner three-dimensional cube are extracted as shown in (Fig. 3). The targets or labels are the entire three-dimensional DNS simulation for the different variables. All the variables are trained simultaneously using a 3D-UNET to map from the inner domain of DNS to the entire DNS cube. We first divide the entire dataset into training (2000), validation(250) and testing(750). The training data is used to compute the maximum (max) and minimum (min) values across all the 2000 samples for each variable. These max-min values are then used to normalize the training, validation and test datasets. 

\par 

The inner cube of the DNS simulation, which corresponds to a size 128 x 128 x 128, is artificially made to 256 x 256 x 256 using three-dimensional cubic interpolation to facilitate it as an input to the 3D-UNET. The 3D U-Net architecture is quite similar to the U-Net in terms of its design.
It consists of both the encoding and decoding components. The encoding and decoding parts have a kernel size of 5 x 5 x 5 with max-pooling corresponding to 2 x 2 x 2. Strides of two in each dimension are used and the model consists of a cumulative 1,794,385 parameters. The learning rate used for training is 0.0001 and the output activation function is 'hyperbolic tangent'. 3D-UNET emerged from the image segmentation community and has shown promising results in emulating DNS simulations.

\par 

We train the 3D-UNET with DNS outputs corresponding to a smaller region as the precursors and the entire domain as the labels. Learning is performed in regression mode using multiple 3D convolutions and deconvolutions. TensorFlow is used for performing the data-driven optimization with the rectified linear units (RELU) employed for non-linear activation. We monitor the validation loss and use early stopping while saving the best model on the validation dataset at train time. The loss is computed as mean squared error and the training is performed iteratively on each sample due to the I/O constraints of the large dataset. The server used for running the jobs is Apollo 6500 Gen10 Plus System comprising of 8 NVIDIA A100 Tensor Core GPUs (40GB). The processor used is Dual AMD EPYC 7402 Processors (24core $@$ 2.8GHz) with 1TB Memory (16 $\times$ 64GB DIMMS) – 3200 MT/s DDR4 Memory. The GPU is able to fit only one sample per training epoch, and hence the choice of batch size is driven by our computational resources. It takes around ~3 days to complete the training task on a single NVIDIA A100 GPU. 
\par 

While the training time may seem large, the benefit of DeepClouds.ai is realized during testing. We were able to generate the test predictions for 750 samples in under ~ 1 hour using a single NVIDIA A100 GPU, typically taking around ~1 day for the DNS multi-CPU simulations. For the testing, central cubes of size 128 x 128 x 128 are extracted from the DNS testing data that DeepClouds.ai has not seen before. Similar to training, the test inputs are first enlarged to a size 256 x 256 x 256 using cubic interpolation and the trained 3D-UNET model performs predictions on these inputs. The test predictions are compared with the outputs from 256 x 256 x 256 discretized cube of the DNS simulations. Thus the 3D-UNET learns to perform three-dimensional spatial enhancement by a factor of 2x.

\section{Results $\textbf{\&}$ Discussion}
\begin{table}
  \centering
  \begin{tabular}{@{}lc@{}}
    \toprule
    Score & Value \\
    \midrule
    Root Mean Squared Error & 5.56e-9 \\
    Peak-Signal-to-noise-ratio & 131 \\
    Linear regression coefficient & 0.67\\
    \bottomrule
  \end{tabular}
  \caption{Various metrics comparing the deep learning-based predictions with the direct numerical simulation outputs.}
  \label{tab:example}
\end{table}
\begin{figure*}
  \centering
  \includegraphics[width=0.8\linewidth]{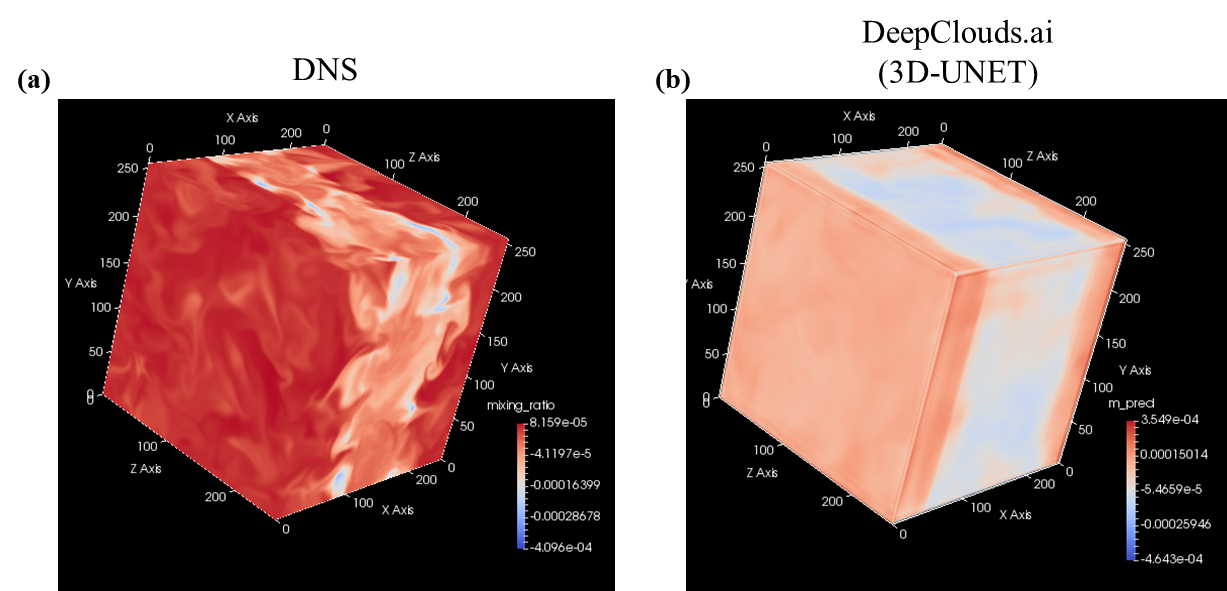}

  \caption{Visualization of vapor mixing ratio ($q_v$) profiles in 3D domain obtained from DNS and 3D-UNET model simulation at t = 420sec on the test dataset}
  \label{fig-4}
\end{figure*}
\begin{figure*}
  \centering
  \includegraphics[width=0.8\linewidth]{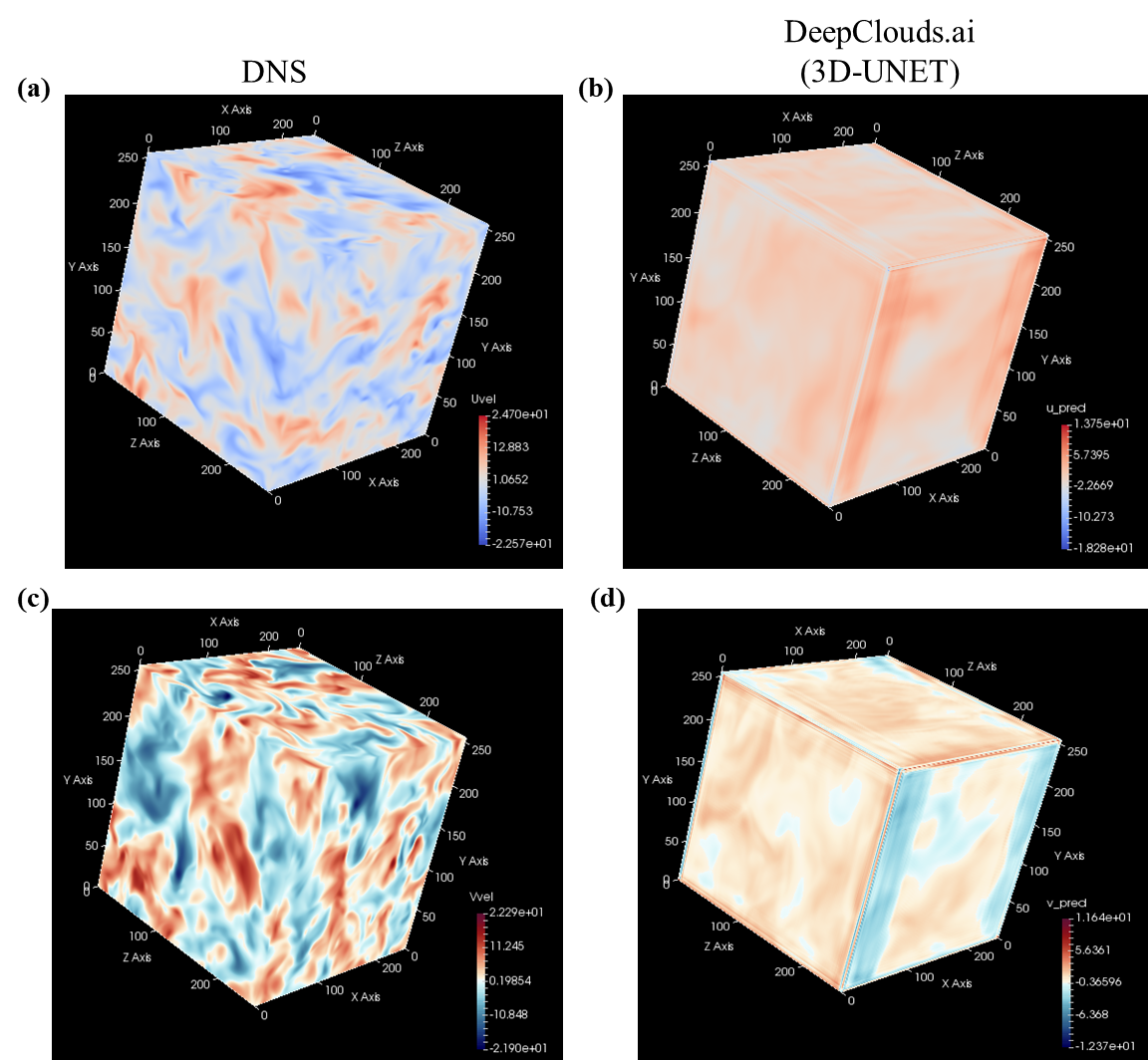}

  \caption{Visualization of (a) u-component and (b) v-component  of 3D velocity profiles obtained from DNS and 3D-UNET model simulation at t = 400sec on the test dataset}
  \label{fig-3}
\end{figure*}

The efficiency of DeepClouds.ai to enlarge the inner 3D region of a cubical cloud DNS simulation for the mixing ratio is shown in Table 2. As compared to the absolute numbers from (Fig. 4), the RMSE is 3-4 orders less than the absolute values of mixing ratio. The data-driven simulations of cloud flows have a high PSNR of 131 dB, as shown by Table 2. The learned deep learning model is trained on DNS data of turbulent flow in an upward increasing cloudy domain, and predictions are made for an unseen test data set.


\par
The vapors mixing ratio ($q_v$) is defined as the mass of water vapor to the mass of dry air. Fig. 4 shows the $q_v$ profiles from DNS (Fig. 4a) and the learned model (Fig. 4b). It is important to note that since the domain is adiabatic, the mixing ratio will be conserved inside the domain. This indicates that when the water vapor content in the flow increases, so does the value of the mixing ratio, and vice versa. The amount of moisture in the flow is critical for following the growth trajectories of cloud droplets. If the water vapor amount in the domain increases with time, it indicates that the cloud droplets are evaporating. Whereas if the water vapor amount in the domain decreases, it is indicative of the expanding droplets by condensation. If this process continues, the droplets will form raindrops, and precipitation will occur. Fig. 4a demonstrates that the mixing ratio is highest around the domain's edges and lower near the center in DNS simulations. DeepClouds.ai (3D-UNET) is able to well capture the same pattern in the predicted $q_v$-profiles (see Fig. 4b). The temperature difference between the flow and the cloud droplets is the cause of high $q_v$ near the boundary. Droplets evaporate as they approach the domain's boundaries, and the mass of water vapor in the flow increases. 

\par

Fig. 3 shows a visual comparison of 3D turbulent flow velocity in streamwise (i.e. u-component) (Fig. 3 (a-b)) and transverse (v-components) (Fig. 3 (c-d)) directions at instantaneous DNS simulation timestep of t= 400 sec. This visualization aims to examine the 3D turbulence intensity within the cloudy domain as it rises upward in time. Fig. 3a shows that the flow profile has a low variance for the u-component relative to the v-component of the 3D velocity profiles from the DNS and 3D-UNET. It is clear that the 3D-UNET performs a reasonable job at representing the smoothness of flow in the streamwise direction.

%

In comparison to streamwise turbulent flow, transverse turbulent flow is substantially rougher (Fig. 3c). This means that the flow in this direction isn't evenly distributed across the domain. The trained model adequately captures the flow's non-uniformity along the v-components (see Fig. 3d). This non-uniformity is caused by the domain's upward movement as well as the existence of cloud droplets within the domain. The surrounding temperature and pressure of the environment change with each time step as the domain rises, resulting in a velocity differential in the turbulent flow. Cloud droplets can travel freely inside the domain, and as a result, they begin to interact with one another. These interactions produce impediments for the flow field, resulting in velocity component non-uniformity. It can be seen from both figures (Fig. 4 and Fig. 3) that our deep learning model performs reasonably well relative to the direct numerical simulation.
%

%
%

%
\section{Conclusions}

In this work, we present DeepClouds.ai which is a data-driven approach to perform direct numerical simulations (DNS) of large domain small-scale processes in the atmosphere. DNS is the only possible option at present to simulate the complex, nonlinear turbulent small-scale flow fields in the atmosphere, particularly that over the physical clouds. It lacks in its ability to be carried out over large domains due to the high computational costs. Even the biggest supercomputer with hundreds of thousands of CPU cores would struggle to perform a DNS over the spatial scales of a real atmospheric cloud. The inability to perform such simulations limits the present numerical weather prediction models in their capacity to form raindrops at small-scales, i.e., the scales at which the processes would occur in reality. Thus, there is a grave need to perform DNS simulations for large domains. 
\par 
DeepClouds.ai is a framework that built using the outputs from DNS simulations enabling the recursive large domain small-scale data-driven simulations. To the best of our knowledge, this is a first attempt towards this step. Our results show realistic simulations of upward rising clear-cloudy turbulent flow. For the deep learning of DNS simulation, we use a 3D-UNET. The datasets and the framework will act as a baseline for benchmarking deep learning enabled direct numerical simulations. DeepClouds.ai serves as a link between the data science community and physicists working on cloud flows. It has the potential to benefit a variety of societal concerns by enabling improved skills in numerical weather prediction systems. 

Future research will focus on using various architectures of computer vision algorithms to improve on the baseline established in this study. For enhanced data-driven DNS, we would additionally incorporate temporal data in addition to spatial data. The datasets and codes will be made publicly available to contribute in the advancement of cloud physics and atmospheric science research. 



\bibliographystyle{unsrt}  
\bibliography{templatePRIME}

\end{document}